\documentstyle[sprocl]{article}

\def\be{\begin{equation}}
\def\ee{\end{equation}}
\def\bea{\begin{eqnarray}}
\def\eea{\end{eqnarray}}

\begin{document}


\title{ON ENERGETICS AND STRUCTURE OF SUB-PARSEC SCALE JETS IN QUASARS} 

\author{M. SIKORA}

\address{N. Copernicus Astronomical Center, Polish Academy of Sciences,
\\ Bartycka 18, 00-716 Warsaw, Poland\\E-mail: sikora@camk.edu.pl} 

\author{G. MADEJSKI}

\address{Stanford Linear Accelerator Center, 
\\Menlo Park, CA 94025, USA\\E-mail: madejski@slac.stanford.edu}


\maketitle\abstracts{In our review of sub-parsec scale jets in quasars, 
we discuss the following issues: observations of parsec and sub-parsec scale 
jets; energy dissipation and particle acceleration; 
radiative processes; magnetic fields, pair content and energetics;
variability and its relation to the central engine activity.
In particular, we describe how internal shocks
can explain properties of gamma-ray flares and  demonstrate that MeV blazars 
(those with luminosity peak
in the 1-30 MeV range) can be unified with GeV blazars (those with
luminosity peak at GeV energies) assuming that 
in GeV blazars the gamma-ray flares are produced via Comptonization of broad 
emission lines,
whereas in the MeV blazars they result from Comptonization of infrared 
radiation of hot dust. We also make predictions about the radiative effects
of bulk Compton process in the soft X--ray band and show how
spectral and variability properties in that band can be used
to constrain structure of jets near their bases.}

\section{Introduction}
Observational data imply that jets in radio-loud quasars
can extend over 8 decades of distance, from milliparsecs to 
hundreds of kiloparsecs.  Such jets are launched in the
vicinity of super-massive black holes and transport energy at a rate
which is sometimes comparable with the accretion luminosity, i.e.
$10^{45}-10^{47}$ erg s$^{-1}$.~\cite{RS,CF} 
Jets are relativistic and therefore Doppler boosting often makes them 
observable only on one side of the quasar. However, the hot spots where 
jets terminate are usually seen on both sides of the central source, 
and emission on both sides shows similar flux.  This suggests that 
the radiation from hot spots is nearly isotropic, and indicates 
that jets are ``light'', i.e. have lower mass densities than the external 
medium. The lightness 
is necessary to explain the nonrelativistic speeds of hot spots
and formation of extended radio lobe structures.~\cite{BC}  
Using the asymmetry of radio brightness of jets and counter-jets, 
Wardle and Aaron~\cite{WA} derived the bulk Lorentz factor 
$\Gamma \sim 3$. Larger Lorentz factors are suggested by X--ray 
observations~\cite{CGC,Tav};  however, because X--ray sensitivity 
is too low to observe counter-jets, such estimates are not direct 
and depend on the details of models of the X--ray production.  

Quasar jets are imaged in the radio band down to parsec scales. There,
they join the central cores, which in turn, due to the synchrotron 
self-absorption, produce very flat spectra, with the energy 
spectral index $\alpha < 0.5$ ($\alpha: F_{\nu} \propto \nu^{-\alpha}$)
(see, e.g., review by Zensus.~\cite{Zensus}  In quasars with jets 
oriented close to the line of sight, the flux of the radio cores 
strongly dominates over that from radio lobes, and the total radio 
spectra are flat.  Because of this, core-dominated radio 
quasars are often called FSRQs (flat spectrum radio quasars), as opposed 
to the lobe dominated radio quasars, which have steep radio spectra 
and are called SSRQs (steep spectrum radio quasars).  Radio cores 
in FSRQs are often variable on monthly time scales, and variability of
the radio flux is often accompanied by an appearance of 
ejecta propagating with apparently superluminal velocities.~\cite{VC,Jor}

The physics of the parsec scale jets (energetics, magnetic fields, 
particle distribution) is determined by the use of images, 
spectra, variability, and polarization properties.  In particular, 
the angle of linear polarization provides information about magnetic 
field orientation relative to the jet axis.  Recent data suggest 
that such orientation changes from being perpendicular to the 
jet closer to the central source to the roughly parallel at somewhat 
larger distances.~\cite{SRH,Lis} This may indicate that at smaller 
distances, the jets are free (ballistic) and that their energy is 
dissipated mainly in perpendicular shocks resulting from collisions 
of the jet inhomogeneities moving with different radial velocities.  
At larger distances, on the other hand, most dissipation is likely to 
come from oblique, reconfinment shocks produced due to interaction of a jet
with external medium.  For a number of objects, there are also 
circular polarization measurements available.~\cite{Hom} 
If circular polarization is produced via the Faraday conversion mechanism, 
it provides information on the number of electrons at mildly 
relativistic energies.  Combined with the information about the total 
power of a jet, this can be used to estimate the pair content.
It should be emphasized that no information about number of mildly
relativistic electrons is available from direct
radio flux observations, because the synchrotron radiation by mildly 
relativistic electrons is self-absorbed.

The parts of jets located close to the cores can be studied also in 
other spectral bands besides radio, despite the fact that they are 
by many orders too compact to be resolved.  In most FSRQs, typical 
thermal components of the quasar emission -- such as the UV radiation 
from an accretion disc, the X--ray radiation from the disc corona, and 
the infrared radiation from dust -- are overshined by the Doppler-boosted 
nonthermal jet radiation.  Variability of this radiation is measured 
on time scales often shorter than those observed in the radio band, 
and this strongly suggests its parsec/sub-parsec origin.  
High polarization in the optical and IR bands 
implies synchrotron mechanism, while the distinct high energy 
components, with luminosity peaks in the MeV -- GeV range,~\cite{Mon} are 
presumably  products of Comptonization of broad emission lines and 
IR radiation of hot dust.~\cite{S94,Bla} If the EC 
(external-Compton) scenario is indeed responsible for the 
high energy components, then observations of the X--ray spectra can 
provide an exceptional opportunity to study the population of the lower energy 
population of the relativistic electrons directly.~\cite{SM}  Furthermore, 
due to  narrowness of the electromagnetic spectrum bands covered by 
dominant broad emission lines and by infrared radiation of hot dust, 
any features such as breaks in the energy distribution of 
electrons should be sharply imprinted in the EC radiation components.
In particular, as it  will be demonstrated below, the shapes of the 
observed high energy spectra are consistent with the two-power-law 
injection function of electrons, and the latter can
result from a two-step acceleration process of electrons as discussed below.

The two component nonthermal spectra are also observed in BL Lac objects, 
which differ from FSRQs as having very weak or undetectable emission lines.  
FSRQs and BL Lac objects form together a class of objects called 
{\it blazars}. It is shown by Fossati {\it et al}.~\cite{Fos} that broadband 
spectra of blazars form a sequence which can be parametrized by their 
total luminosities.  In this sequence, FSRQs are the most luminous objects. 
Both their low and high energy spectral components appear to be least 
extended to the high energies, and their $\gamma$--ray luminosities 
during flares strongly dominate over synchrotron luminosities.  
The least luminous blazars are represented by the X--ray selected BL Lac 
objects. Their synchrotron spectra extend up to hard X--rays, 
and the $\gamma$--ray spectra reach TeV energies. $\gamma$--ray luminosities 
in the TeV-emitting BL Lac objects usually do not dominate over 
synchrotron luminosities.  Low luminosity  BL Lac objects are probably 
associated with nuclei of radio galaxies accreting at a low accretion rate, 
and $\gamma$--rays are very likely produced by the SSC 
(synchrotron-self-Compton) process.  
In the more luminous, radio selected BL Lac objects the EC process can be 
dominant,~\cite{Mad} in similarity to FSRQs.

In this presentation we focus on jets in quasars, where the interaction 
of the jet with external radiation is significant and the presence or lack
of radiative effects of this interaction can be used to determine physical 
parameters and constraints on structure of jets on sub-parsec scales.  
This paper is organized as follows: in \S2, we analyze the 
energetics of intrinsic collisions of material in the jet;  
in \S3, we present the motivation for introduction of the two-step 
stochastic acceleration of electrons;  in \S4, we discuss the formation 
of spectral breaks in the high energy spectral components; in \S5, \S6, 
and \S7, respectively, we derive the magnetic field intensity, the pair 
content, and the average power of a jet.  In \S8, we formulate our 
predictions regarding the production of soft X--ray precursors via 
Comptonization of the external radiation by the cold electrons in a jet, 
and list the main conclusions in \S9.  

\section{Energy dissipation}
Cooling time scales of ultrarelativistic electrons, which 
in quasar jets  produce 
synchrotron radiation in the optical band and inverse-Compton radiation
in the MeV -- GeV range, is much shorter than the dynamical/propagation 
time scale. Therefore, such electrons must be accelerated {\it in situ},
in sites where the jet loses part of its  energy and the shocks are formed. 
The jet energy can be dissipated in 
internal shocks -- formed in collisions between inhomogeneities 
in a jet~\cite{S94,Spada} or following reconnection of magnetic 
fields,~\cite{RL} -- and/or in external shocks, formed via 
reconfinment of the jet by external medium~\cite{San,KF} or due to
collisions of a jet with external clouds~\cite{DC}.  Perpendicular 
orientation of magnetic fields to the jet axis, inferred from the 
polarization measurements in the optical band~\cite{Imp} and at the 
high radio frequencies~\cite{Lis} favors the scenario where at 
sub-parsec distances, the particles are accelerated in transverse shocks 
which are likely to be produced via collisions of inhomogeneities flowing 
down the jet with different radial velocities.  A given collision event 
is then responsible for a given flare, and the time scale of such a collision, 
as measured in the comoving frame of the shocked plasma, is
\be t_{coll}' \simeq t_{fl} {\cal D}  \, ,\ee
where $t_{fl}$ is the observed time scale of the flare, and
\be {\cal D} \equiv {1 \over \Gamma (1 - \beta \cos{\theta_{obs}})} \ee
is the Doppler factor of the shocked plasma.
The amount of energy dissipated during the collision is 
\be E_{diss}' \simeq 
{L_{fl}' t_{coll}' \over \eta_{rad} \eta_e} \simeq
{ L_{fl} t_{fl} \over f {\cal D}^3 \eta_{rad} \eta_e} \, ,  \ee
where $\eta_e$ is the fraction of dissipated energy used to accelerate
electrons, $\eta_{rad}$ is the (average) radiative efficiency of electrons,
and
\be L_{fl} \equiv 4 \pi \left(\partial L_{fl} \over \partial \Omega \right)
=  4 \pi \left(\partial L_{fl}' \over \partial \Omega' \right) {\cal D}^4
= L_{fl}' f {\cal D}^4 \ee
is the apparent luminosity. All primed quantities are as measured in 
the source comoving frame. The factor $f$ expresses the additional dependence 
of the apparent luminosity on the observed angle, if the intrinsic radiation 
is anisotropic. For the EC process,
$f\simeq ({\cal D}/\Gamma)^2$, while for synchrotron and 
SSC radiation $f\simeq 1$.~\cite{Der}

In the above relations it was assumed that during a collision,  
the shocked plasma moves with a constant bulk Lorentz factor.
In general this is not true: depending on the initial parameters of 
inhomogeneities, the shock can accelerate or decelerate, and 
during the collision the shock structure can evolve from a double 
one to a single shock (forward or reverse).  However, if inhomogeneities 
are intrinsically identical, then the forward-reverse shock structure 
is symmetric in the shocked plasma frame (= discontinuity contact 
surface frame), and the bulk Lorentz factor of the shock is constant.
This case is adopted in our quantitative analysis, not only for
simplicity, but also because it provides the highest efficiency
of energy dissipation.

Assuming that the inertia of inhomogeneities is dominated by protons
(i.e. $n_e/n_p \ll m_p/m_e$), and noting that 
\be E_{diss} \equiv \eta_{diss}( E_1 + E_2) = \Gamma E_{diss}' \, , \ee
where $E_{1} \simeq  (N_p/2) \Gamma_{1} m_p c^2$ 
and $E_{2} \simeq  (N_p/2) \Gamma_{2} m_p c^2$ are energies of
inhomogeneities, we find that total number of protons involved in 
the collision is 
\be N_p \simeq { 2\Gamma E_{diss}' \over 
\eta_{diss} (\Gamma_1 + \Gamma_2) m_p c^2} \simeq 
{2 L_{fl} t_{fl} \over \kappa f {\cal D}^3 \eta_{rad} \eta_e m_pc^2} \, ,
 \ee
where 
\be \kappa \equiv{ E_{diss}' \over N_p m_p c^2} =  
{\eta_{diss} (\Gamma_1 + \Gamma_2) \over 2 \Gamma} \ee
is amount of energy dissipated per proton in units of $m_pc^2$.
For $\Gamma_2 >\Gamma_1 \gg 1$, energy and momentum conservations give
\be \Gamma \simeq \sqrt {\Gamma_1 \Gamma_2} \ee 
and
\be \eta_{diss} \simeq 
{((\Gamma_2/\Gamma_1)^{1/2} - 1)^2 \over (\Gamma_2/\Gamma_1)+1} \, ,\ee
and then
\be \kappa \simeq
{ ((\Gamma_2/\Gamma_1)^{1/2} - 1)^2 \over 2(\Gamma_2/\Gamma_1)^{1/2}} \, . 
\label{eq:kappa} \ee
For  $\Gamma_2/\Gamma_1 = 2.5$, which corresponds with 
$\eta_{diss} \simeq 0.1$, $\kappa \simeq 0.11$.

\section{Particle acceleration}
As it was demonstrated by Sikora and Madejski,~\cite{SM} energy flux
in powerful jets in quasars cannot be dominated by pair plasma.
This is because such jets would produce much larger fluxes of 
soft X--ray radiation than is observed.  Hence, we assume that the 
inertia of jets is dominated by protons. In this case, the structure 
of shocks and the structure of the turbulence around them are both 
determined by protons. Provided that the magnetic fields are amplified
up to equipartition with protons, the protons are effectively
accelerated both by 1st order and by 2nd order Fermi process.~\cite{Jones}
Those which  reach energies $> 10^9$ GeV, interact efficiently with 
ambient photons and trigger (mainly via photo-meson process) the 
synchrotron-supported pair cascades.~\cite{MB} However, such a model 
fails to reproduce very hard X--ray spectra of FSRQs.~\cite{SM2} This 
may indicate that too few protons reach sufficiently high energies to 
power pair cascades.  

Alternative possibility is that the high energy radiation in FSRQs is 
produced by directly accelerated electrons.  However, for the 
acceleration of electrons to occur via the Fermi process, these electrons 
must be first pre\-hea\-ted\-/pre\-acce\-lera\-ted up to energies $\gamma_b$, at 
which point, the magnetic rigidity of electrons becomes comparable with
magnetic rigidity of thermal protons, i.e. when their momenta are equal:
\be m_e \sqrt{\gamma_b^2 -1} \simeq  m_p \sqrt {\gamma_{p,th}^2-1} \, ,
\label{eq:gb} \ee
where 
\be \gamma_{p,th} -1 = \eta_{p,th} \kappa  \label{eq:pth} \ee
is the average thermal proton energy in the shocked
plasma, and $\eta_{p,th}$ is the fraction of the dissipated energy 
tapped to heat the protons. For the reasonable assumption that 
$\Gamma_2/\Gamma_1 <10$, $\gamma_{p,th}-1 < 1$, i.e. the thermal
proton plasma is at most mildly relativistic.

Noting that the average energy of injected electrons is
\be \bar\gamma_{inj}-1 \simeq {n_p m_p  \over n_e m_e} \eta_e\kappa =
{n_p m_p\over n_e m_e}{\eta_e \over \eta_{p,th}} (\gamma_{p,th} -1) \, , \ee
we find that for  $\gamma_b$ and $\bar\gamma_{inj} \gg 1$
\be {\gamma_b \over \bar\gamma_{inj}} \sim 
{n_e \over n_p}{\eta_{p,th} \over \eta_e} \, 
\left({\gamma_{p,th}+1 \over \gamma_{p,th}-1}\right)^{1/2} \, .\ee
Hence, for non- or mildly relativistic shocks and for 
$\eta_{p,th} \sim \eta_e$, the threshold energy for the Fermi 
acceleration of electrons significantly exceeds the average 
energy of electrons, even if $n_e=n_p$.  That implies that the 
often-considered bulk preheating process is not able to provide 
an adequate number of electrons with $\gamma \ge \gamma_b$.  
But, as it was recently demonstrated in numerical PIC (particle-in-cell) 
simulations, the preheating/preacceleration mechanism has a stochastic 
character and a significant fraction of electrons can reach such 
energies.~\cite{Diec,SH}  Furthermore, the fact that in FSRQs, the 
X--ray spectral index $\alpha_X$ is never negative, implies 
that most of electrons occupy the lowest energies, i.e. the median 
Lorentz factor is $\ll \bar\gamma_{inj}$.  Therefore, 
this scenario does not predict a thermal bump in the preaccelerated electron 
energy distribution.  Instead, the preacceleration mechanism, in 
similarity to the Fermi process, produces a power-law energy distribution 
of electrons.  It should be emphasized here that because the X--rays 
in FSRQs are produced in the slow cooling regime (see next Section), 
electrons with low energies are not populated by cooling effect, 
but are directly injected with such energies.

\section{Electromagnetic spectra}
The basic feature of the high energy spectra in FSRQs --- 
a spectral break between the X--ray and the $\gamma$--ray bands ---
has a natural explanation in terms of the EC model.  In this model,
X--ray spectra are produced by electrons with radiative cooling time 
scale $t_{cool}'$, longer than the collision time scale $t_{coll}'$
({\it slow cooling regime}), whereas $\gamma$--rays are produced by 
electrons with $t_{cool}' < t_{coll}'$ ({\it fast cooling regime}).  
Noting that the angle averaged cooling rate  of electrons, dominated by 
Comptonization of external radiation, is 
\be \vert \dot \gamma \vert \simeq {  \sigma_T  \over m_e c} u_{ext}' \gamma^2 \, , \ee
we obtain that the angle averaged cooling time scale is
\be t_{cool}' \simeq {\gamma \over \vert \dot \gamma \vert} \simeq
{m_e c \over   \sigma_T}{1 \over  \gamma u_{ext}'}  \, , 
\label{eq:tc} \ee
where $u_{ext}'$ is the energy density of an external radiation field.
Then, from $t_{cool}' = t_{coll}'$, where $t_{coll}'$ is given by Eq. (1),
the break  in the electron distribution is at energy
\be \gamma_c \simeq {m_e c \over  \sigma_T} 
{1 \over u_{ext}' t_{fl} {\cal D}} \, . \label{eq:gac} \ee
For $\gamma < \gamma_c$, the slope of the electron distribution is
the same as the slope of the injection function; for 
$\gamma > \gamma_c$, the slope of the electron energy distribution is 
steeper by $\Delta s=1$
($s: N_{\gamma} \propto \gamma^{-s}$).

Since time scales of flares in FSRQs are rarely shorter than 1 day, 
the distances from the central source where they are produced, 
\be r_{fl} \sim (r_{fl} /\Delta r_{coll}) c t_{fl} {\cal D} \Gamma \, ,\ee
are expected to be larger than 0.1 parsec. At such distances,
contribution to $u_{ext}'$ is dominated  by the diffuse 
components of the broad emission line light and  
of the infrared radiation of hot dust. Noting that 
\be u_{ext}' = {1 \over c} \int I_{ext}' d\Omega' =
{1 \over c} \int I_{ext} {\cal D}_{in}^{-2} d\Omega \simeq 
u_{diff} \Gamma^2  \,  \ee
where 
\be {\cal D}_{in} = {1 \over \Gamma (1-\beta \cos \theta_{in})} \, \ee
and $\theta_{in}$ is the angle between the photon direction and the jet axis,
we predict  that  the break in an electron energy distribution at $\gamma_c$ 
should be imprinted in the electromagnetic spectrum at frequency
\be \nu_{EC,c} \simeq {\cal D}^2 \gamma_c^2 \nu_{diff}
\simeq \left({m_e c^2 \over \sigma_T}\right)^2 
{\nu_{diff} \over u_{diff}^2 t_{fl}^2 \Gamma^4 }
\label{eq:nc} \ee 
and that the spectrum should change the slope around $\nu_{EC,c}$ by 
$\Delta \alpha_{X,\gamma} \simeq 0.5$.  For typical energy 
densities expected in the BELR (broad emission line region) 
as well as that provided by the hot dust radiation field, the break
is predicted to be located in the range of $10^{20} - 10^{22}$ Hz (Sikora
{\it et al}.),  in agreement with observations.~\cite{Mon,Fos} 

In most FSRQs, during high energy flares the slope of $\gamma$--ray 
spectra measured by EGRET~\cite{Pohl} is $\alpha_{\gamma} \le 1$, 
while the X--ray spectra show slopes $\alpha_X \simeq 
0.6-0.7$.~\cite{Com,Don}  The smaller than the predicted value 
$0.5$ of $\Delta \alpha_{X,\gamma}$ can be explained by taking into
account contribution to the soft and mid-X--ray bands of the SSC
component~\cite{IT,Kubo}.  More challenging are the so called ``MeV 
blazars.'' In these objects, the spectra peak in the $300$ keV - $30$ MeV 
range and the $\gamma$--ray spectra are very soft ($\alpha_{\gamma}>1.4$), 
while the X--ray spectra are very hard ($\alpha_X \le 0.5$).~\cite{Tav} 
The values of $\Delta \alpha_{X,\gamma} \ge 1.0$  measured 
in those sources are much too large to be explained 
solely by the effect of cooling, and instead, it is necessary to
postulate an additional break in the electron energy distribution 
besides $\gamma_c$.  It is tempting to identify such a break with the 
break at $\gamma_b$, predicted by the two-step acceleration process discussed 
in the Section 3.  External photons scattered by electrons with 
energies $\gamma_b$, are boosted to energies
\be \nu_{EC,b} \simeq  {\cal D}^2 \gamma_b^2 \nu_{diff} \, . \ee
For $2.5 <  \Gamma_2/\Gamma_1 < 10$ and $\eta_{p,th} \sim 0.5$, 
the break in the electron injection function is 
$600 < \gamma_b < 1700$ (see Eqs.~\ref{eq:kappa}-\ref{eq:pth} and Table 1), 
and should appear in the electromagnetic 
spectrum at around $1$ GeV if $\nu_{diff} =\nu_{BEL}$, and at lower
energies (by a factor of $\sim 10 - 30$) if $\nu_{diff} =\nu_{IR}$.
In the former case $\nu_{EC,b} \gg \nu_{EC,c}$ and the $\gamma$--ray spectra
in the EGRET range should have slopes $\alpha_{\gamma} \le 1$, while 
in the latter case $\nu_{EC,b} \sim \nu_{EC,c} \le 30$ MeV, and the 
$\gamma$--ray spectra in the EGRET range should be soft.  Hence, we 
conclude that for blazars producing most of their radiation in the 
GeV band (and hereafter called GeV blazars) the $\gamma$--ray flux 
is probably produced at smaller distances (closer to the central source), 
where the production of $\gamma$--rays is dominated by Comptonization 
of broad emission lines, while the MeV blazar phenomenon is likely 
to be produced at larger distances, where production of $\gamma$--rays 
is dominated by the Comptonization of infrared radiation from dust.  

The above scheme can also explain two other differences between 
MeV blazars and GeV blazars.  One, already mentioned above, is that 
in MeV blazars the X--ray spectra should be much harder than in GeV blazars. 
This can result from the lower contribution of the SSC component to 
the X--ray band in MeV blazars as compared to GeV blazars 
(Sikora {\it et al}., in preparation). The other difference is that 
in the spectra of MeV blazars, in contrast to GeV blazars, the 
thermal UV bump  is quite prominent.~\cite{Tav}  This difference can be 
explained noting that magnetic fields are weaker at larger distances, 
and therefore in MeV blazars the synchrotron spectra are shifted to 
lower frequencies, revealing the UV bump.

\section{Magnetic fields}
 
Electrons with energies from the range $[\gamma;\gamma+d\gamma]$
produce synchrotron radiation at the rate
\be L_{\nu_{syn}'}d\nu_{syn}'  \sim 
(N_{\gamma}d\gamma) \vert \dot \gamma_{syn}\vert m_e c^2
\sim (N_{\gamma}d\gamma) c \sigma_T \gamma^2 u_B' \ee
and the angle averaged EC radiation  at the rate
\be L_{\nu_{EC}'}d\nu_{EC}' \sim 
( N_{\gamma}d\gamma) \vert \dot \gamma_{EC}\vert m_e c^2
\sim (N_{\gamma}d\gamma) c \sigma_T \gamma^2 u_{ext}' \ee
where
\be \nu_{syn}' \sim (2e/3\pi m_e c) \gamma^2 B' \ee
\be \nu_{EC}' \sim \gamma^2 \nu_{ext}' \ee
and $u_B' = {B'}^2/8\pi$ is the magnetic energy density.
Noting that $L_{\nu_{syn}}d\nu_{syn}  = L_{\nu_{syn}'}d\nu_{syn}' 
{\cal D}^4$ and 
$L_{\nu_{EC}}d\nu_{EC} 
=  L_{\nu_{EC}'}d\nu_{EC}' ({\cal D}^6/\Gamma^2) $
we find that the ratio of synchrotron to EC luminosity, both 
produced by the same population of electrons in the same energy 
range, radiating in the optically thin regime, is
\be {\cal R}_{syn/EC} \equiv 
{\nu_{syn} L_{\nu_{syn}} \over \nu_{EC} L_{\nu_{EC}}} \sim
{u_B' \over u_{ext}'} \left(\Gamma  \over {\cal D} \right)^2 \, .
\label{eq:R} \ee
From  Eqs.~(\ref{eq:gac}) and (\ref{eq:nc}), we have 
\be u_{ext}' \simeq  {m_e c \over   \sigma_T}
{1 \over t_{fl} (\nu_{EC,c}/\nu_{diff})^{1/2} } \ee
and inserting this into Eq.~(\ref{eq:R}), we obtain 
\be 
u_B' \simeq u_{ext}'{\cal R}_{syn/EC} ({\cal D}/\Gamma)^2  \simeq
{m_e c \over \sigma_T}
{{\cal R}_{syn/EC} ({\cal D}/\Gamma)^2 \over 
t_{fl} (\nu_{EC,c}/\nu_{diff})^{1/2} } \, .
\label{eq:ub} \ee
For $t_{fl} \sim 3$ day, ${\cal R}_{syn/EC} = 0.1$, $\nu_{EC,c} = 10$ MeV
and $\nu_{diff} = 10$ eV, and the observer located at 
$\theta_{obs} = 1/\Gamma$, Eq.~(\ref{eq:ub}) gives 
$B' \sim  \sqrt {8 \pi u_B'} \sim 0.6$ Gauss. 

We can estimate the energy density of the relativistic electrons and 
determine whether they are in equipartition with magnetic fields.
Relations $E_{rad}' = \eta_{rad} \eta_e E_{diss}'$
and $E_e' = (1-\eta_{rad}) \eta_e E_{diss}'$ give
\be E_e' = {1-\eta_{rad} \over \eta_{rad}} E_{rad}' \, , \ee
where
\be E_{rad}' \sim L_{fl}' t_{coll}' \sim 
{L_{fl} t_{fl} \over {\cal D}^3 ({\cal D}/\Gamma)^2 } \, . \ee
Hence, energy density of the relativistic electrons is
\be u_e' = {E_e' \over \pi a^2 \Delta_{sh}'} \simeq
{1-\eta_{rad} \over \eta_{rad}} \, 
{L_{fl} \over 2 \pi c^3 (a/a_{\Gamma})^2 t_{fl}^2 g 
{\cal D}^6({\cal D}/\Gamma)^2} \, , \label{eq:ue} \ee
where  $a \le a_{\Gamma} = c t_{fl} {\cal D}$ is the cross-sectional 
dimension (radius) of a shocked plasma sheet, and 
$\Delta_{sh}'= 2 c t_{fl} {\cal D} g$ is the thickness of the sheet.  
The scaling factor  $g$, which relates the thickness of the shocked
plasma sheet to the time scale of the flare/collision, depends 
on $\Gamma_2/\Gamma_1$ and on the adiabatic index. The analytical form
of $g$ is derived and presented in Moderski {\it et al}. (in preparation), 
and the example values of $g$ are shown in the Table 1.  

Combining Eqs.~(\ref{eq:ub}) and (\ref{eq:ue}), we obtain
\be {u_B' \over u_e'} \simeq
{ 2 \pi m_e c^4 \over \sigma_T}{\eta_{rad} \over 1- \eta_{rad}} \,
{g (a/a_{\Gamma})^2 {\cal R}_{syn/EC}   \over (\nu_{EC,c}/\nu_{diff})^{1/2}}
{t_{fl} \over L_{fl}}\, 
 {\cal D}^6 ({\cal D}/\Gamma)^4 \, . \ee
The equation demonstrates very strong dependence of  $u_B' / u_e'$ 
 on $\Gamma$ and on the observed angle $\theta_{obs}$. 
Provided that blazars  are observed at 
$\theta_{obs} \sim 1/\Gamma$,
the equipartition condition, $u_B' \sim u_e'$, implies
\be \Gamma  \simeq \left( {\sigma_T \over 2 \pi m_e c^4} 
{ (\nu_{EC,c}/\nu_{diff})^{1/2} \over 
{\cal R}_{syn/EC} \, g \, (a/a_{\Gamma})^2}\, {1-\eta_{rad} \over \eta_{rad}}
\, {L_{fl} \over t_{fl}} \right)^{1/6} \, .
\label{eq:Gamma} \ee
For flares with luminosities $L_{fl} \sim 10^{48-49}$ergs s$^{-1}$ and 
time scales 
$t_{fl} \sim 3$d, and assuming
$a=a_{\Gamma}$, $\Gamma_2/\Gamma_1 =2.5$, and $\eta_{rad}=0.5$, 
we obtain $\Gamma \sim 18-26$.

\begin{table}[t]
\caption{Dependence of the internal shock parameters on $\Gamma_2/\Gamma_1$.
Values of $g_0$ and $g$ are calculated for the adiabatic index of
the shocked plasma $\hat \gamma = 5/3$. Analytical formulas for 
 $g_0$ and $g$ are presented in 
Moderski {\it et al}. (in preparation).\label{tab:exp}}
\vspace{0.2cm}
\begin{center}
\footnotesize
\begin{tabular}{|c|c|c|c|c|c|c|c|c|}
\hline
{$\Gamma_2/\Gamma_1$} & {$\eta_{diss}$} & {$\kappa$} & 
{$\gamma_{p,th}-1$} & {$\gamma_b$} & {$\bar\gamma_{inj}$} & {$n_+/n_p$}
& {$g_0$} & {$g$}\\ 
\hline
$1.5$ & $0.020$ & $0.020$ & $0.010$ & $265$ & $15.3$ & $0.2$ & $0.273$ & 
$0.067$ \\
$2.5$ & $0.097$ & $0.107$ & $0.053$ & $609$ & $18.0$ & $2.4$ & $0.641$ & 
$0.150$ \\
$5.0$ & $0.255$ & $0.342$ & $0.171$ & $1120$ & $20.4$ & $7.6$ & $1.237$ & 
$0.255$ \\
$10.0$ & $0.425$ & $0.739$ & $0.370$ & $1722$ & $22.2$ & $15.5$ & $2.025$ & 
$0.346$\\
\hline
\end{tabular}
\end{center}
\end{table} 

\section{Pair content}

The pair content is
\be {n_{+} \over n_p} = ({n_e \over n_p} -1)/2 \ee
where
\be {n_e \over n_p} = \eta_e {m_p \kappa \over m_e (\bar \gamma_{inj}-1)} \ee
\be \bar \gamma_{inj} = { \int_1^{\gamma_{max}} \gamma Q \, d\gamma
\over \int_1^{\gamma_{max}} Q \, d\gamma} \ee
and $Q$ is the electron injection function.
For a double power-law injection function,
$Q \propto \gamma^{-p_L} $ for $1 <\gamma < \gamma_b$, and 
$Q \propto \gamma^{-p_H}$ for $\gamma_b < \gamma < \gamma_{max}$,
and observationally  suggested indices,  $1 <p_L < 2$ and $p_H > 2$,
\be \bar \gamma_{inj} \simeq 
{ (p_L -1) (p_H-p_L) \over (2-p_L)(p_H-2)} \gamma_b^{2 - p_L} \, . \ee
For $p_L = 1.8$  ($\to \alpha_X = (p_L -1)/2 \sim 0.45$ and $\alpha_{\gamma} =
p_L/2 \sim 0.9$ in the EGRET band  of GeV blazars), and $p_H = 2.8$
($\to \alpha_{\gamma} = p_H/2 = 1.4$ in the EGRET band of MeV blazars),  
and $2.5 < \Gamma_2/\Gamma_1 < 10$, the number of pairs per proton
is $2.4 < n_+/n_p < 15.5$ (see Table 1). 

\section{The power of a jet}
Energy carried by the cold inhomogeneities prior to the collision is
(see Eq. 6)
\be E_i \simeq {N_p \over 2} m_p c^2 \Gamma_i \simeq
{L_{fl} t_{fl} \over \kappa f {\cal D}^3 \eta_{rad} \eta_e} \Gamma_i
\, ,\ee
where $i=1,2$.
The time it takes to eject the inhomogeneities from the central source is
\be t_{ej,i} \simeq {\lambda_0 \over c \beta_i \Gamma_i}
\simeq {{\cal D}  \over \Gamma_i} \, g_0 t_{fl} \ee
where $\lambda_0 = g_0 c {\cal D} t_{fl} $ is the
proper length of an inhomogeneity (the thickness of a shell segment),
and $g_0$ is the factor depending on $\Gamma_2/\Gamma_1$
and on the adiabatic index of the shocked plasma (see Table 1).
The time-averaged power of a jet is
\be P_j \sim \Psi {E_1 + E_2 \over t_{ej,1} + t_{ej,2}} \simeq
\Psi {N_p m_p c^3 \Gamma^2 \over \lambda_0} \simeq
\Psi {L_{fl} \over 2 \eta_e \eta_{rad} \kappa g_0}
{\Gamma^2 \over f {\cal D}^4} \, , \label{eq:Pj} \ee
where $\Psi$ is the duty cycle factor.
For $L_{fl} \sim 10^{48}$erg s$^{-1}$, ${\cal D} \sim \Gamma \sim 15$,
$\Gamma_2/\Gamma_1 \sim 2.5$, $\eta_e \sim \eta_{rad} \sim 0.5$ and
$\Psi \sim 0.1$, Eq.~(\ref{eq:Pj}) gives  
$P_j \sim 3 \times 10^{46}$~erg~s$^{-1}$.

\section{X--ray precursors}
Prior to the collision, electrons in inhomogeneities are cold, but being 
pulled by streaming protons they  scatter external photons
and produce the beamed bulk-Compton radiation.~\cite{BS,SM}  
The spectrum of this radiation peaks around
\be \nu_i \simeq {\cal D}_i \Gamma_i \nu_{diff} \ee
and has luminosity  
\be L_{X,i} 
\simeq  \delta N_{e,i} \vert\dot\epsilon_{e,i}\vert {\cal D}_i^4 \, , \ee
where
\be \delta N_{e,i} = (N_e/2) {\rm Min} [1;r_{BC}/\lambda_0 {\cal D}_i] \, ,
\label{eq:Min} \ee
\be N_e  \simeq
{E_{e,inj}' \over  \bar \gamma_{inj} m_e c^2} \simeq
{\eta_e E_{diss}' \over \bar \gamma_{inj} m_e c^2} \simeq
{ L_{fl} t_{fl} \over  f {\cal D}^3 \eta_{rad} \bar \gamma_{inj} m_e c^2} \, ,
\ee 
\be \vert\dot\epsilon_{e,i}\vert \simeq \Gamma_i^2 c \sigma_T u_{ext} \, ,\ee
\be {\cal D}_i \equiv {1 \over \Gamma_i (1 - \beta_i \cos{\theta_{obs}})} \, ,
\ee
$r_{BC}$ is the distance within which most of the bulk Compton radiation 
is produced.  
The factor  Min$[1;r_{BC}/\lambda_0 {\cal D}_i]$ in  Eq.~(\ref{eq:Min}) 
expresses the fact that in general, due to light travel effects 
and the finite length of the source, not all electrons 
contribute to the radiation observed at a given instant.   

Bulk-Compton flares from cold inhomogeneities are expected to precede 
the nonthermal flares by 
\be \delta t_i \sim {r_{fl} \over c \Gamma_i {\cal D}_i} \simeq
 {{\cal D} \Gamma \over {\cal D}_i \Gamma_i} t_{fl} 
{\Delta r_{coll} \over r_{fl}} \, .  \ee
Above formulae are very approximate, because they do not take into
account that the observed bulk Compton radiation at any given instant
includes contributions from a finite distance range over which 
$u_{ext}$ can change significantly.  However, making the conservative 
assumption that the jet is developed (accelerated and collimated) 
not earlier than at a distance where electron energy losses are already  
dominated by Compton interactions with the broad emission line flux 
rather than the direct radiation from an accretion disc, we can calculate 
the minimum bulk Compton luminosity, using an approximation that 
$u_{ext} \simeq u_{BEL} \sim$ const for 
$r < r_{BEL}$ and 0 for $r > r_{BEL}$.
Noting that $r_{BEL} \sim 3 \times 10^{17} \sqrt {L_{UV,46}}$ cm
and  $u_{BEL} \sim 0.03$ erg cm$^{-3}$ (Peterson~\cite{Pet}),
we find that for $\theta_{obs} = 1/\Gamma= 1/15$, $\Gamma_2/\Gamma_1 = 2.5$,
$L_{fl} = 10^{48}$~erg~s$^{-1}$ and $t_{fl}=3$ day, the 
precursor produced by the faster-moving inhomogeneity should have 
luminosity $L_{X,2} \sim 3 \times 10^{46}$ ergs s$^{-1}$; should peak
at around h$\nu_2 \simeq 3.2$ keV; and should precede the nonthermal flare
by $\delta t_2 \simeq 0.7 t_{fl}$.  The precursor produced by
the slower moving inhomogeneity  would have 
luminosity $L_{X,1} \sim 5 \times 10^{45}$ erg s$^{-1}$; would peak
around h$\nu_1 \simeq 1.8$ keV; and would precede the nonthermal flare
by $\delta t_1 \simeq 1.8 t_{fl}$.
Since  typical soft X--ray luminosities are of the order 
$10^{46}$~erg~s$^{-1}$, we can infer from the above estimates that 
jets propagating through much denser quasi-isotropic radiation fields 
than would be provided by the BELR are predicted to produce much stronger 
soft/mid X--ray radiation than is observed.  This puts severe constrains 
on the jet structure at $10^2-10^3$ gravitational radii from the BH, 
where quasi-isotropic radiation is strongly dominated by the 
very dense direct radiation of an accretion disc and its corona.
Hence, the lack of very large soft X--ray excesses suggests that
jets near their bases are very wide (meaning that they have relatively 
large opening angle) and/or are slower than at larger distances.

\section{Summary}
Multi-wavelength observations of blazars provide a valuable tool 
allowing us to learn about the structure and physical properties 
of sub-parsec-scale jets in quasars.  The observations suggest:

\noindent 
$\bullet$ Rapid flares and transverse magnetic fields inferred
from polarization data $\to$
{\it Streams of matter in small-scale jets are unsteady, and 
collisions between inhomogeneities propagating down the jet 
with different radial velocities can easily form transverse shocks};

\noindent
$\bullet$ Large powers of jets and relatively low soft/mid X--ray 
luminosities $\to$ {\it Inertia of jets is dominated by protons and the number
of pairs per proton is $<15$};

\noindent
$\bullet$ Very hard X--ray spectra $\to$ {\it X--rays and $\gamma$--rays are 
produced by electrons accelerated directly, just as is the case 
for the synchrotron radiation, rather than being injected following 
the proton-induced pair cascades};

\noindent
$\bullet$ Differences between properties of  GeV  and MeV blazars
$\to$ {\it  EC model can explain these differences 
assuming a two-power-law
electron injection  function. The break in the injection function can be 
related to the threshold energy for the Fermi acceleration of electrons};

\noindent
$\bullet$ Luminous, short term  $\gamma$--ray flares in FSRQs $\to$ {\it
magnetic fields are much below  equipartition with the relativistic plasma,
unless $\Gamma \ge 18-26$};

\noindent
$\bullet$ No prominent spectral signatures of  bulk Compton process in
the soft/mid X--ray bands $\to$ {\it Collimation and/or acceleration of jets 
occurs beyond $10^2-10^3$ gravitational radii from the central source}.

\section*{Acknowledgments}
This work was partially supported by Polish KBN grant 5 P03D 002 21
and NASA Chandra observing grant to Stanford University.  
M.S. thanks the Stanford Linear Accelerator Center
for its hospitality during the completion of this work, 
and the Korean Institute for Advanced Study for hosting this interesting
meeting. 


\end{document}